\documentclass[prl,twocolumn,amsmath,amssymb,aps,superscriptaddress,groupedaddress,nofootinbib,showpacs]{revtex4}
\usepackage{graphicx}
\usepackage[usenames, dvipsnames]{color}
\usepackage{tabularx}
\usepackage{siunitx}
\usepackage{xcolor}
\usepackage[
verbose,
colorlinks=true,
naturalnames=true,
linkcolor=blue,
citecolor=blue,
]{hyperref}

\begin{document}

\preprint{APS/123-QED}

\title{Sub-atomic constraints on the Kerr geometry of GW150914}

\author{Julian Westerweck}
\affiliation{Albert-Einstein-Institut, Max-Planck-Institut f{\"u}r Gravitationsphysik, \\ Callinstra{\ss}e 38, 30167 Hannover, Germany}
\affiliation{Leibniz Universit{\"a}t Hannover, 30167 Hannover, Germany}

\author{Yotam Sherf}
\affiliation{Department of Physics, Ben-Gurion University, Beer-Sheva 84105, Israel}

\author{Collin D. Capano}
\affiliation{Albert-Einstein-Institut, Max-Planck-Institut f{\"u}r Gravitationsphysik, \\ Callinstra{\ss}e 38, 30167 Hannover, Germany}
\affiliation{Leibniz Universit{\"a}t Hannover, 30167 Hannover, Germany}
\affiliation{Department of Mathematics, University of Massachusetts, Dartmouth, MA 02747, USA}

\author{Ram Brustein}
\affiliation{Department of Physics, Ben-Gurion University, Beer-Sheva 84105, Israel}

\date{\today}

\begin{abstract}
    We obtain stringent constraints on near-horizon deviations of a black hole from the Kerr geometry by performing a long-duration Bayesian analysis of the gravitational-wave data immediately following GW150914. GW150914 was caused by a binary system that merged to form a final compact object. We parameterize deviations of this object from a Kerr black hole by modifying its boundary conditions from full absorption to full reflection, thereby modeling it as a horizonless ultracompact object. Such modifications result in the emission of long-lived monochromatic quasinormal modes after the merger. These modes would extract energy on the order of a few solar masses from the final object, making them observable by LIGO. By putting bounds on the existence of these modes, we show that the Kerr geometry is not modified down to distances as small as $5.8 \times 10^{-19}$ meters away from the horizon. Our results indicate that the post-merger object formed by GW150914 is a black hole that is well described by the Kerr geometry.
\end{abstract}

\maketitle

\section{Introduction}
General relativity (GR) predicts the existence of black holes which possess a horizon, a surface that acts as a perfect absorber.
The exterior vacuum geometry of stationary rotating black holes in GR is that of the Kerr geometry \cite{Kerr:1963ud}.

A binary black hole merger results in a rotating, perturbed black hole which then relaxes to equilibrium by emitting gravitational waves (GWs) at specific frequencies, the frequencies of its ringdown or quasinormal modes (QNMs). In GR, the spectrum of the QNMs is completely determined by the black hole mass and spin. Previous QNM analyses of the GW ringdown from binary black hole mergers have yielded broad consistency with the remnant being a Kerr black hole~\cite{TheLIGOScientific:2016src, Isi:2019aib, LIGOScientific:2020tif, Capano:2021etf, LIGOScientific:2021sio}. The first overtone of the dominant QNM was found in LIGO's GW150914 event by Ref.~\cite{Isi:2019aib} (and in other events in Ref.~\cite{LIGOScientific:2020tif, LIGOScientific:2021sio}). In Ref.~\cite{Capano:2021etf}, a subdominant fundamental mode was found in GW190521. In all cases, the recovered modes were consistent with GR.

Here we present a method for testing the validity of the Kerr geometry down to microscopic distances away from the horizon, in the region where gravity becomes strong, and apply it to the LIGO GW150914 data.
In contrast to a black hole, a horizonless object is not a perfect absorber of GWs, and could be distinguished from a black hole by its post-merger GW emission.
As the interaction of the GWs with the interior matter of the object is expected to be weak, the infalling waves could propagate into the object and re-emerge after some time delay \cite{Brustein:2018ixz}.
Thus, to model a horizonless object, we modify the boundary conditions to allow GW reflection at a surface that is a relative distance $\epsilon \ll 1$ away from the would be horizon (see Eq.~(\ref{edef})).
This description is equivalent to the wave passing through the interior and leads to the same functional dependence \cite{Vilenkin:1978uc}.

Imposing boundary conditions in the Kerr geometry that allow reflection near the horizon leads to the appearance of additional QNMs. The initial ringdown modes are very similar to those of a Kerr black hole, as they result from excitations of the photon sphere.
The additional modes are long-lived, nearly monochromatic GWs, expected to appear after a time delay and dominate the emission at times long after the merger (see for example Fig. 3 of \cite{Cardoso:2014sna}). Their frequency is proportional to the rotational frequency of the black hole, while their lifetime is $\tau \sim M |\ln \epsilon|^2$, where $M$ is the black hole mass ($\ln = \log_e$) \cite{Starobinsky:1973aij,Vilenkin:1978uc,Maggio:2018ivz}. For GW150914, the frequency of such modes would be $\sim \SI{210}{\hertz}$ --- well within LIGO's sensitive band --- with lifetimes in the range $\SI{30}{\second} \lesssim \tau \lesssim \SI{8000}{\second}$ (assuming mass and spin estimates from \cite{Nitz:2019hdf}, and allowing $\epsilon \in [10^{-45}, 10^{-5}]$).

The amplitude of the additional modes is determined by the total energy falling in through the initially formed trapped surface \cite{Gupta:2018znn,Mark:2017dnq}.
Since about the same amount of energy falls into the trapped surface as is emitted during the merger, \cite{Gupta:2018znn,Mark:2017dnq}, we expect that the same amount will be channeled to the additional QNMs. In GW150914, we estimate that the total amount of extracted energy is $\sim 3\,\text{M}_\odot$ [see Eq.~\eqref{ee}] and therefore should be detectable with high signal-to-noise ratio (SNR). The additional signal is weak but extremely long lived. By using a long integration time we can place stringent constraints on $\epsilon$.

In this work, we directly constrain $\epsilon$ by performing a long-duration Bayesian analysis of the GW150914 post-merger data.
We develop new parameter estimation methods to overcome the challenges posed by the long duration of the signal and analysed data. Through these, we can probe the near-horizon region of a rotating black hole with unprecedented accuracy, and constrain its geometry down to microscopic distances away from the horizon.

The additional ringdown modes have some resemblance to the so-called black hole echoes \cite{Cardoso:2016rao,Cardoso:2016oxy}, in that they are associated with reflection from the black hole and that they produce a long-duration post-merger GW signal. However, the additional modes differ in some significant aspects from echoes. In the echoes model, the initial merger signal repeats itself at regular intervals, with a decay rate that is treated as a free parameter. The model has five free parameters in total. In our model, the resulting GW signal is a damped sinusoid which resembles in form the standard black hole ringdown modes. The frequency, decay time, and amplitude are all determined by the modified boundary conditions at the reflecting surface, and the mass and angular momentum of the black hole.

Several echoes searches were performed in \cite{Abedi:2016hgu,Ashton:2016xff,Conklin:2017lwb, Westerweck:2017hus,Nielsen:2018lkf,Uchikata:2019frs,LIGOScientific:2020tif}. While some of the searches reported  evidence for near-horizon structure \cite{Abedi:2016hgu,Conklin:2017lwb}, others \cite{Uchikata:2019frs,Ashton:2016xff,Westerweck:2017hus,Nielsen:2018lkf} found low statistical evidence for echoes. An extended search that uses the model proposed in \cite{LIGOScientific:2020tif} was done using the LIGO-Virgo gravitational-waves transient-catalog-2 (GWTC-2) for 31 black hole events. That search reported no statistically significant evidence for echoes in the data. Some implicit constraints on $\epsilon$ can be deduced from the null results of these searches \cite{Ashton:2016xff,Westerweck:2017hus,Nielsen:2018lkf,Cardoso:2019rvt}.
However, these constraints depend on several uncertain modelling assumptions.

Previous efforts to constrain $\epsilon$ using electromagnetic emission from black holes were based on the idea that if the horizon of a black hole is replaced by a hard surface at a fractional distance $\epsilon$ away from the horizon, the electromagnetic emission from such a surface can be observed and could be used to place limits on the luminosity of black holes \cite{Lu:2017vdx}. Several analyses \cite{Doeleman:2012zc,EventHorizonTelescope:2019dse,EventHorizonTelescope:2019ggy} eventually led to impressive nominal results $\epsilon\lesssim 10^{-16}$ \cite{Zulianello:2020cmx,Lu:2017vdx} (corresponding to a distance of $\sim \SI{e-6}{\meter}$).
However, obtaining concrete limits using this method requires making many assumptions \cite{Lu:2017vdx}, including about the surrounding matter. For additional discussions of the caveats and limitations of this method, see \cite{Lu:2017vdx,Carballo-Rubio:2018jzw}.

Fortunately, assuming that the Einstein equivalence principle holds, the dynamics of GWs are only sensitive to the geometry, and the interaction between GWs and matter is extremely weak, and therefore independent of specific environmental models.
This allows us to obtain extremely strong constraints: we find $\epsilon < \SI{3.3e-24}{}$ (90\%-credible interval), which corresponds to a distance between the reflective surface and the Kerr horizon of no more than \SI{5.8e-19}{\meter} in the Boyer-Lindquist coordinate distance.

\section{Theoretical framework}
\label{Theory}

The invariant line-element of a Kerr black hole in Boyer-Lindquist coordinates is
\begin{equation}
    \begin{split}
        ds^2~=&~-\left(1-\dfrac{2Mr}{\Sigma}\right)dt^2-\dfrac{4Mr}{\Sigma}a \sin^2 \theta d\phi dt+\dfrac{\Sigma}{\Delta}dr^2+\\&\Sigma d\theta^2 ~+\left((r^2+a^2)\sin^2 \theta +\dfrac{2 Mr}{\Sigma} a^2 \sin^4 \theta\right)d \phi^2 ~.
    \end{split}
\end{equation}
Here $a$ is the spin parameter, $\Sigma=r^2+a^2\cos^2 \theta$, and $\Delta=r^2+a^2-2Mr=(r-r_+)(r-r_-)$, with $r_\pm=M\pm \sqrt{M^2-a^2}$. The angular velocity of the horizon, $\Omega$, is related to $a$ through $\Omega=(a/M)/2r_+=\chi/2r_+$, with the dimensionless spin parameter $\chi=a/M$.

Gravitational perturbations in the exterior vacuum Kerr geometry obey the Teukolsky equations \cite{Teukolsky:1972my,Teukolsky:1973ha}, which reduce to an eigenvalue problem when regularity of the solution is imposed. The resulting radial equation can be simplified by changing variables \cite{Detweiler:1977gy} and using tortoise coordinates $dr_{*}/dr=(r^2+a^2)/\Delta$, taking the final form
\begin{gather}
    \dfrac{d^2 {}_s\!\Psi_{lm}}{dr^{2}_{*}}-V(r,\omega) {}_s\!\Psi_{lm}~=~0~.
    \label{wavef}
\end{gather}
For gravitational perturbations, the spin is $s=\pm 2$.
In tortoise coordinates, the spatial coordinates are Euclidean and hence Eq.~(\ref{wavef}) describes potential scattering in flat space. The expression for the effective potential $V(r,\omega)$ can be found in \cite{Detweiler:1977gy}.

We find the spectrum of the additional QNMs by imposing boundary conditions at infinity and at the near-horizon surface $r_{NH}$, which is at a relative distance $\epsilon$ above $r_+$,
\begin{equation}
    \epsilon=\dfrac{r_{NH}-r_+}{r_+}.
    \label{edef}
\end{equation}
The solutions of Eq.~(\ref{wavef}) behave approximately as follows,
\begin{align}
    & \Psi ~\sim ~e^{i \omega r_{*}},   & r_{*}&\rightarrow \infty, \\
    &\Psi~\sim ~ e^{-i \omega r_{*}}+\mathcal{R}e^{i \omega r_{*}},
    &r_{*}&\rightarrow r_{*}(r_{NH}),
\end{align}
where $\mathcal{R}$ is the reflection coefficient of the surface, and the complex frequency $\omega=\omega_R+i\omega_I$ has to satisfy Eq.~(\ref{wavef}). The real and imaginary part of $\omega$ are related to the frequency $f$ and damping time $\tau$ of the QNM by $\omega_{R} = 2 \pi f$ and $\omega_{I}^{-1}=\tau$.
An additional unknown phase accounts for the propagation through the interior and is absorbed into the phase $\phi$ in the waveform of Eq.~(\ref{ht}), while we marginalise over the phase of the signal in the numerical analysis.

For a Kerr black hole, the reflection coefficient is zero at the horizon. We modify the boundary conditions at $r=r_{NH}$ such that $\mathcal{R}$ is nonvanishing. In general, $\mathcal{R}$ may depend on the frequency. However, since we consider only a small frequency range $M|\omega_R-m\Omega|\ll 1$, we take $\mathcal{R}$ to be a constant.

We choose a perfectly reflecting boundary condition, $\mathcal{R}=1$. This choice is justified on grounds that if the Einstein equivalence principle holds for the interaction of GWs with the black hole, then the object's surface can only either be fully absorbing ($\mathcal{R} \ll 1$), or fully reflecting ($1-\mathcal{R} \ll 1$). Partial absorption ($0 < \mathcal{R} < 1$) would require the object to contain a membrane or other viscous fluid capable of dissipating GWs~\cite{Thorne:1986iy,Yunes:2016jcc}. However, such models only yield non-negligible absorption when unknown exotic matter is considered~\cite{Yunes:2016jcc,Sherf:2021ppp}. Heuristically, if the matter is not exotic, then the absorption through the object's surface scales  as $1/\tau$. This means that the deviation from total reflection should scale as $r_+/\tau\ll 1$, which means that $1-\mathcal{R}\ll 1$. Conversely, firewall and fuzzball models yield almost full absorption due to the large density of black hole microstates and the small energy gaps between them~\cite{Mathur:2005zp,Guo:2017jmi}. This makes them functionally indistinguishable from classical GR black holes. We therefore focus on the pure reflection case and fix $\mathcal{R} = 1$.
A more detailed argument is found in the Appendix.

For perfect reflection and $s=-2$, the solution for the dominant contribution $l=2$ can be found analytically \cite{Starobinsky:1973aij,Vilenkin:1978uc} (also see \cite{Maggio:2018ivz}), yielding
\begin{align}
\omega_R &\simeq~m\Omega \pm \dfrac{\pi}{2 |r^0_{*}|}\left( \nu + 1 \right),
    \label{rw}\\
\omega_{I} &\simeq~\dfrac{2M\left(\omega_R-m\Omega\right)r_+}{225|r_*^0|\left(r_+-r_-\right)} \left[\omega_R(r_+-r_-)\right]^{5}.
    \label{iw}
\end{align}
Here, $|r^0_{*}|\sim\int dr \sqrt{g_{rr}}\sim M\left(1+(1-\chi^2)^{-1/2}\right)|\ln\epsilon|$; we choose the dominant overtone number $\nu=1$ (not to be confused with the QNM-overtone number $n$). The remaining modes have an almost identical frequency and are practically indistinguishable from the $\nu=1$ mode. Furthermore, the amount of energy stored in the higher overtones $\nu\ge 2$ is expected to be much lower than that stored in the dominant mode.

The solutions contain two types of signals, damped or superradiant for a positive or negative  sign of $\omega_R(\omega_R-m\Omega)$, respectively. \cite{Starobinsky:1973aij,Vilenkin:1978uc}
Only two absolute values of $\omega_R$ appear for each value of $|m|$, as changing the sign of both $m$ and the second term in~Eq.(\ref{rw}) in turn only changes the sign of $\omega_R$.

Alternatively, the damping properties of the modes can be explained from an interior perspective where, similar to \cite{Brustein:2017nis}, the scattering cross-section of the outgoing waves is positive and leads to a damped rather than amplified waveform, see \cite{Starobinsky:1973aij} and Appendix for further details. As noted in \cite{Brustein:2017nis,Brustein:2018ixz}, a heuristic description is that the would-be BH is effectively in an excited state and it decays to equilibrium with a lifetime $\tau$.

We focus on the case $\epsilon\ll 1$ such that $|\ln \epsilon|\gg 1$.
Then Eq.~(\ref{rw}) is mostly governed by the angular frequency of the object $\omega_R\approx \chi/r_+$ and Eq.~(\ref{iw}) corresponds to a large damping time $\tau \sim r_+ |\ln \epsilon|^2$.
The large damping time allows us to constrain $\epsilon$ by analyzing a long duration of post-merger data.

\section{Signal model}

Our signal model reflects the  damped oscillatory properties of the modes, and relies on the knowledge of the initial merger phases from which we can extract all other parameters of the black hole. We then assume a smooth transition between the early to late time phases \cite{Price:1994pm,Buonanno:2006ui}.

We use a quasi-normal mode to model the late-time post merger signal,
\begin{align}
    (h_{+} + i h_{\times}) (t) &= {}_{-2} S_{lm} (\iota, \varphi) A e^{-t/\tau} e^{i (2 \pi f t + \phi)} \Theta(t-t_0)
    \label{ht} ~,
\end{align}
which is parametrized by five intrinsic parameters.
These are the amplitude $A$, frequency $f = \omega_R / 2 \pi$, damping time $\tau = \omega_I^{-1}$ and initial phase $\phi$ of the damped sinusoid, and a start time $t_0$ of the signal.
If the prompt QNM emission occurs at $t=0$, then $t_0$ describes the time delay between this and the start of the additional QNM signal.
The spin-weighted spheroidal harmonics ${}_{-2}S_{lm}$  depend on the inclination $\iota$ and azimuth angle $\varphi$.
Here, we consider the dominant spherical mode $l = m = 2$ and approximate the spheroidal harmonics by spin-weighted spherical harmonics \cite{Berti:2007zu,Berti:2005gp}.
For $\epsilon \ll 1$, the frequency $\omega_R$ in Eq.~(\ref{rw}) is governed by the object's angular velocity,
\begin{align}
    M \omega_R &= \frac{\chi}{ \left( 1 + \sqrt{1 - \chi^2} \right)} + \frac{\pi \sqrt{1 - \chi^2}}{ \left| \ln \epsilon \right| \left( 1 + \sqrt{1 - \chi^2} \right)}.
    \label{freq_w}
\end{align}
For a set of example parameters compatible with GW150914, $M \approx 62 \text{M}_\odot, \chi \approx 0.67$, and for $\epsilon = 10^{-25}$, we would find $M \omega_R \approx 0.4$ and $f \approx \SI{211}{\Hz}$.
This range of parameters guarantees the validity of Eq.(\ref{iw}), since as pointed out in \cite{Starobinsky:1973aij}, the derivation relies on the assumption that $M\omega_R < 1$, $a \omega_R< 1$ and $M(\omega_R-m\Omega)\sim \frac{1}{|\ln \epsilon|} \ll  1$.

The amplitude $A$ is determined by the  total  energy  (and  angular  momentum)  that is carried away by the GWs to infinity (see \cite{Brustein:2018ixz}).
The total emitted energy is determined at the merger \cite{Gupta:2018znn}, we label it by $\Delta E=E_{init}+E_{rot}$.
Then, by using the non-relativistic approximation, such that $E_{rot}=\frac{1}{2}E_{init} \Omega^2r_+^2$, we find
\begin{gather}
    \Delta E~=~E_{init}\left(1+\dfrac{ \chi^2}{8}\right).
    \label{ee}
\end{gather}
For the same example parameters, rotational effects lead to a correction of the emitted energy by an increase of $\sim 5\%$ compared to the non-spinning case, yielding $\Delta E \approx 3.2 M_\odot$.
In the superradiant case in contrast, most of the rotational energy is extracted by the emitted GW, such that $E_{rot}\sim M \Omega^2 r_+^2\sim 5 M_{\odot}$.

For the final black hole of GW150914, the majority of the energy falling in and being reflected passes through the effective potential barrier, while only a small part is then reflected back in, leading to weak further pulses.
Approximating the peak of the Kerr BH effective potential barrier through the WBK-method~\cite{Seidel:1989bp} to first order, we find $V_{\text{max}}\approx (M\omega_{\text{QNM}})^2$, where $\omega_{\text{QNM}}$ is the fundamental Kerr QNM's frequency, $M\omega_{\text{QNM}} \sim 0.5$. As $\omega_R \lesssim \omega_{\text{QNM}}$, the outgoing wave mostly passes the potential barrier.

To calculate the amplitude, we evaluate the emitted energy $\Delta E$ by using the leading order GW flux formula,
\begin{gather}
    \dot{E}_{GW}~=~\dfrac{D_L^2}{32\pi}\int \langle \dot{h}_{\mu\nu}\dot{h}^{\mu\nu}\rangle d\Omega~.
    \label{quadru}
\end{gather}
Here the dot denotes a time derivative,  $D_L$ is the luminosity distance, $h_{\mu\nu}$ is  the  waveform  in  the transverse–traceless gauge, $d\Omega$ is an element of solid angle, and angular brackets denote averaging over short wavelengths.
We approximate the integral in Eq.~(\ref{quadru}) by noticing that the emitted GWs are approximately monochromatic with $\omega_R\simeq 2\Omega$, yielding $\dot{E}_{GW}\approx \frac{1}{4} D_L^2\langle |\dot{h}|^2\rangle$.
Then, by taking $h(t)$ from Eq.~(\ref{ht}) and for  $\epsilon\ll 1$ such that $\omega_R \tau \gg1 $,
the final expression for the amplitude becomes
\begin{align}
    A &= \dfrac{4}{\omega_R D_L  }\left(\dfrac{\Delta E}{\tau}\right)^{1/2}.
    \label{AE}
\end{align}
In Eq.~(\eqref{AE}) the parameters $\omega_R$, $\tau$ and $\Delta E$ are given in Eqs.~\eqref{rw}, \eqref{iw}, and \eqref{ee}, respectively.
The explicit form of $\omega_R$ is given in Eq.~(\ref{freq_w}), while for $\tau$ it is
\begin{align}
    \tau~&=~\dfrac{225M}{32\pi}\left(\dfrac{1+\sqrt{1-\chi^2}}{\sqrt{1-\chi^2}}\right)^6\dfrac{|\ln \epsilon|^7}{\left(\chi |\ln \epsilon|+\pi\sqrt{1-\chi^2}\right)^5}
    \label{tau}
\end{align}

We fix the parameter $t_0$ to an arbitrary value some time after the merger.
To prevent contamination of the analysis from the standard ringdown modes, we choose a time that is large compared to the lifetime of these modes, but short compared to the lifetime of the additional signal, $t_0 = \SI{32}{\second}$. Because the amount of energy emitted during this relatively short time is small and because the SNR is determined by the total collected energy, we do not lose much diagnostic power by this choice.
As the damping time increases for smaller $\epsilon$, this approximation is more accurate for the expected small values of $\epsilon$.

In addition to $\epsilon$, the parameters varied in the analysis are right ascension $\alpha$, declination $\delta$, polarisation $\psi$, inclination $\iota$, luminosity distance $D_L$, final mass $M$, final spin $\chi$, and energy radiated in the primary GW emission, $\Delta E$. Equations~(\ref{freq_w}), (\ref{ee}), (\ref{AE}) and (\ref{tau}) then determine the parameters of the damped sinusoid template. The phase $\phi$ of the signal is marginalised over analytically. We use as priors for the source parameters the posteriors found in \cite{Nitz:2019hdf}, calculating $M$, $\chi$ and $\Delta E$ from the component parameters via fitting formulae to numerical relativity \cite{Hofmann:2016yih, Tichy:2008du, lalsuite}. For the only additional parameter of our model, $\epsilon$, we use a log-uniform prior in the interval $-45\le \log_{10} \epsilon \le -5$.

We use Bayesian methods to estimate the signal parameters from the data. The toolkit \texttt{PyCBC Inference} \cite{pycbcgithub, Biwer:2018osg} is used to compute the likelihood and estimate the posterior probability distributions. The parameter space is sampled using the parallel-tempered Markov-chain Monte Carlo sampler \texttt{emcee\_pt} \cite{ForemanMackey:2012ig,Vousden:2015}.

We modify the standard parameter estimation analysis to prevent influences from boundary effects. The expected signal persists for a longer time than the currently manageable duration of the analysis. We therefore need to restrict the time series data to a shorter time window, which introduces a discontinuity from the sharp cut-off at the window edges.
This leads to artefacts in the frequency domain response function of the whitening filter. To avoid this, we remove the times containing these artefacts, and we employ a heterodyning procedure to reduce the computational cost of generating long template waveforms (see Appendix).

\section{Results}

\begin{figure}
        \includegraphics[width=\columnwidth]{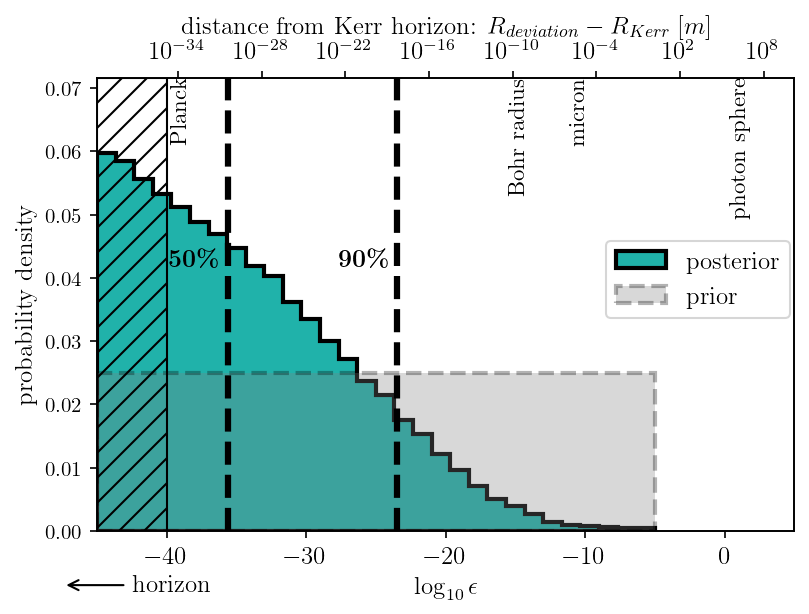}
    \caption{The histogram shows the marginal posterior for the fractional deviation from the Kerr geometry, $\epsilon$, measured for the proposed signal for GW150914.
        The prior for $\log_{10} \epsilon$ is flat, as shown in the shaded region.
        The dashed lines mark the one-sided 50th and 90th percentile upper bound.
        On the top axis the coordinate distance between reflective surface and horizon corresponding to $\epsilon$ is shown for the post-merger black hole in GW150914, and hatching indicates distances below the Planck length.
        As the distance posterior is virtually identical to the posterior for $\log_{10} \epsilon$, we only show the latter and use the maximum likelihood values for mass and spin from \cite{Nitz:2019hdf} to convert from $\log_{10} \epsilon$ to the distance scale.
    }
    \label{fig:epsilon_posterior}
\end{figure}

\begin{figure}
    \includegraphics[width=\columnwidth]{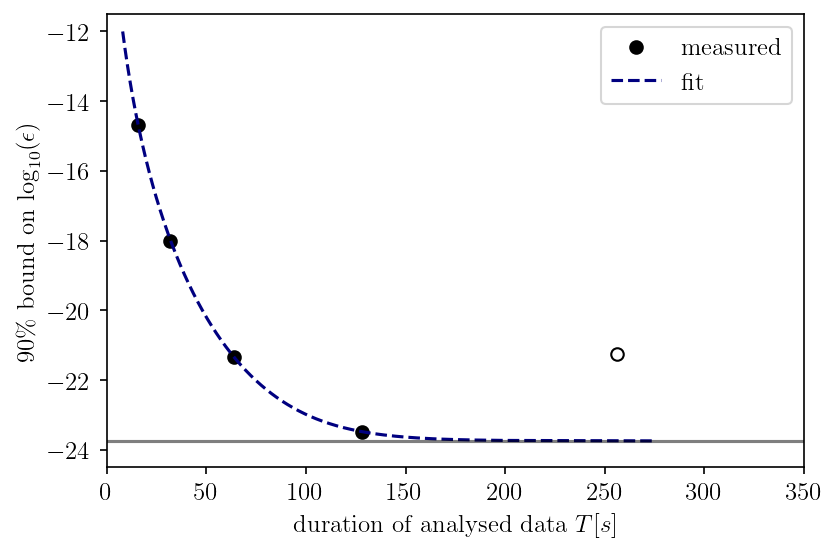}
    \caption{Bounds placed on $\epsilon$ for the analysis of different durations of data.
        A curve of the form $T(|\ln \epsilon|) \sim c |\ln(\epsilon)|^2 \ln\left(1 - \left(a-{b}/|\ln(\epsilon)| \right)^2 \right)$, with constants $a,b,c$, is fitted to the data (see Appendix). The fit asymptotically approaches $\log_{10} \epsilon \approx -23.7$. For longer segments of data, the bound increases again (empty circle), as the posterior begins to be dominated by lines in the power spectral density of the noise.
    }
    \label{fig:epsilon_trend}
\end{figure}

Applying our analysis to \SI{128}{\second} of data starting \SI{32}{\second} after GW150914 yields the posterior on $\epsilon$ shown in Fig.~\ref{fig:epsilon_posterior}. The results are consistent with the absence of the searched signal, as the posterior peaks toward the lower boundary. Our upper bound on the 90\%-credible interval is $\log_{10} \epsilon = -23.5$. For the post-merger black hole of GW150914, this bound corresponds to a distance between the reflective surface and the Kerr event horizon of no more than \SI{5.8e-19}{\meter} in the Boyer-Lindquist coordinate distance.

To validate this result, we repeat the analysis on data before GW150914, when no signal is expected, as well as on Gaussian noise. We also inject a simulated signal with $\log_{10} \epsilon = -21$ into detector noise to verify that the analysis can detect a louder signal when present. We find that the posterior on $\epsilon$ does peak toward the injected value in the latter case, whereas in noise the posterior and limits are similar to what we obtain for the GW150914 post-merger data (see Appendix).

To investigate the effect of the amount of time analyzed on the bound on $\epsilon$, we repeat the analysis using time segments \SI{16}{\second}, \SI{32}{\second}, \SI{64}{\second}, and \SI{256}{\second}.
The results are shown in Fig.~\ref{fig:epsilon_trend}. As expected, the upper bound on $\epsilon$ increases as we analyze shorter time segments than the \SI{128}{\second} we use above. This suggests that analyzing longer times would yield even better limits. However, in the \SI{256}{\second} analysis the bound is worse than what we obtain with \SI{128}{\second}. This is because the posterior on $\epsilon$ begins to be dominated by lines in the power spectral density of the noise as the analysis time increases, leading to weaker constraints. Overcoming this would require removing lines from the data, which is outside the scope of this work.

Using the \SI{16}{\second} to \SI{256}{\second} results (for which lines are not an issue) we estimate the best limit that could theoretically be obtained with GW150914. Fitting the expected relationship between observation time and $\epsilon$,
we find that the best 90\%-constraint using arbitrary lengths of data would be  $\log_{10} \epsilon \approx -23.7$. This limit arises due to a combination of the SNR of GW150914 and the energy available in this system that could be converted to the long duration QNMs. Since the potential signal is a damped sinusoid, the recoverable SNR asymptotes to a fixed value for infinite observation time. This in turn puts a limitation on the smallest $\epsilon$ that can be measured. As can be seen in Fig.~\ref{fig:epsilon_trend}, we are close to this limit with the \SI{128}{\second} analysis time.

\section{Conclusion and outlook}

We performed the first long-duration QNM analysis of the post-merger data of GW150914,
and ruled out the existence of long-lived additional QNMs.
Through this, we put a bound on the validity of the Kerr geometry down to fractional distances from the horizon as small as $\epsilon < \SI{3.3e-24}{}$, which is equivalent to a coordinate distance $< \SI{5.8e-19}{\meter}$.
Our result improves existing bounds by many orders of magnitude and indicates that the GW150914 post-merger object is a black hole that is well described by the Kerr geometry.

Based on the fit in Fig.~\ref{fig:epsilon_trend}, we conclude that to significantly improve our bounds will require a black hole merger with larger SNR than GW150914.

By combining results over multiple events, and with improving sensitivity of future detectors, it should be possible to eventually probe spacetime geometry down to Planck scales above the horizon. This could provide confirmation of the Kerr nature of astrophysical black holes all the way to their horizons.

\section{Acknowledgements}
The authors thank Ofek Birnholtz, Alex B. Nielsen, and Reinhard Prix for valuable discussions, and Paolo Pani and Enrico Barausse for helpful comments.
This work benefited from discussions at the ``Gravitational wave searches and parameter
estimation in the era of detections" workshop held January 12-18, 2020 at Ringberg castle, Tegernsee lake. Calculations were performed on the Atlas computer cluster of the Albert Einstein Institute Hannover. The research of RB and YS was supported by the Israel Science Foundation grant no. 1294/16. The research of YS was supported by the Negev scholarship.

This research has made use of data obtained from the Gravitational Wave Open Science Center (https://www.gw-openscience.org/ ), a service of LIGO Laboratory, the LIGO Scientific Collaboration and the Virgo Collaboration. LIGO Laboratory and Advanced LIGO are funded by the United States National Science Foundation (NSF) who also gratefully acknowledge the Science and Technology Facilities Council (STFC) of the United Kingdom, the Max-Planck-Society (MPS), and the State of Niedersachsen/Germany for support of the construction of Advanced LIGO and construction and operation of the GEO600 detector. Additional support for Advanced LIGO was provided by the Australian Research Council. Virgo is funded, through the European Gravitational Observatory (EGO), by the French Centre National de Recherche Scientifique (CNRS), the Italian Istituto Nazionale di Fisica Nucleare (INFN) and the Dutch Nikhef, with contributions by institutions from Belgium, Germany, Greece, Hungary, Ireland, Japan, Monaco, Poland, Portugal, Spain.

\bibliography{main.bib}

\section{Appendix}
\subsection{Justifying the assumption of full reflection}
Here we elaborate on the arguments given in the main text and provide further explanations for justifying full reflection.

One can understand, heuristically, the scaling of $\omega_R$ and $\tau$, the frequency and decay time of the additional modes. In tortoise coordinates, the near horizon geometry looks flat and Eq.~(\ref{wavef}) can be viewed in terms of a wave propagating in a cavity of length $r_+|\ln\epsilon|$. The scaling of the decay time $\tau$ can be understood in terms of ideas that were introduced in \cite{Brustein:2017nis} and elaborated on in \cite{Brustein:2018ixz}. We briefly review them here and refer the reader to the original articles for further details.

First recall from Eq.~(\ref{rw}) that  the ``proper" angular frequency of the additional modes is $\;\omega_R \sim \frac{1}{r_+|\ln{\epsilon|}}\;$. This means that a co-rotating GR external observer would view them as having a wavelength
$\;\lambda\sim r_+ |\ln{\epsilon}|\;$.  The source of the GWs is the ultracompact object which has an area of about $A\sim M r_+$. The transmission cross-section for such long wavelength modes for an area $A$ is proportional to the ratio $A/\lambda^2$, which scales as $Mr_+/\lambda^2\sim\frac{1}{|\ln{\epsilon}|^2}\;$.
The decay time is inversely proportional to the transmission rate, so scales as  $\;\tau\sim |\ln{\epsilon}|^2\;$.  The scaling $A/\lambda^2$ results from the assumption that the gravitational force acts equally on all forms of matter according to the Einstein equivalence principle.

The heuristic argument that we have just reviewed can also be applied to the case of imperfect reflection at the surface $r_{NH}=r_+(1+\epsilon)$.  Such scenarios require exotic matter which in some cases may violate fundamental principles \cite{Sherf:2021ppp} and are therefore disfavoured. In the case that the reflection is not parametrically small, a case which corresponds to nearly full absorption and so, effectively,  to a horizon, the  mode's decay time would scale as it does for the case of total reflection. The key point  is that partial absorption occurs at the surface $r_{NH}$. Then, the absorption through this surface would scale as $\;A/\lambda^2\sim\frac{1}{|\ln{\epsilon}|^2}\;$. When the angular momentum of the GW is taken into account, one finds that the absorption through the object's outer  surface  scales precisely as $1/\tau$. This means that the deviation from total reflection should scale similarly. Consequently, $1-\mathcal{R}\ll 1$ since $r_+/\tau \ll 1$.

In the majority of echo models, the reflection coefficient is an arbitrary constant that is put by hand; see \cite{Chen:2020htz} and reference therein. None of the reviewed models elaborate on the underlying mechanism that provides the absorption properties of the would-be black hole. Many of them refer to the fundamental papers that motivate horizon scale corrections, such as the firewall and fuzzballs proposals. However, a closer look reveals that a partial absorption of GWs that is comparable to black hole absorption is an unrealistic situation that is not compatible with fundamental physical properties.

For example, in the firewalls-inspired models and the fuzzball proposal \cite{Mathur:2005zp,Guo:2017jmi}, due to the large entropy and density of states and the small energy gap between the black-hole microstates, an infalling quantum is almost fully absorbed. Fuzzball absorption is therefore almost identical to the black hole absorption (see  \cite{Tyukov:2017uig} for specific examples). In \cite{Wang:2019rcf} it was argued that (Eq.~3), for $\omega \ll T_H$, with the Hawking temperature $T_H$,
\begin{equation}
\mathcal{R} = \exp(-\omega/T_H) (\gamma \omega)^{-\omega/T_H}.
\end{equation}
This means that $\mathcal{R}=1$ to exponential accuracy, or
\begin{equation}
\mathcal{R} = 1- \omega/T_H.
\end{equation}
The Hawking temperature $T_H$ in natural units is $1/r_+$, so the intrinsic frequencies that we discuss obey this condition. Similarly in \cite{Cardoso:2019apo}, they argue that $\mathcal{R}$ is close to one, except for special frequencies that correspond to the intrinsic frequencies of the quantum black hole, which are of order $1/R_S$, where $R_S=2M$ is the Schwarzschild radius.

If one wishes to model the object's absorption by an alternative dissipation mechanism as in the membrane paradigm, one needs to assume the existence of an exotic matter.
To show this, it is possible to model the object's intrinsic dissipation in terms of its effective viscosity as in the membrane paradigm \cite{Thorne:1986iy}.
In \cite{Sherf:2021ppp} it is shown that the absorption coefficient $\gamma_{abs}$ scales as $\gamma_{abs}\sim \eta/\eta_{BH}$, where $\eta_{BH}$ is the BH viscosity. The absorption is negligible for all known matter forms. For example, a highly viscous cold neutron star has $\gamma_{abs}\sim 10^{-8}$, while non-rotating strongly magnetized neutron stars and fictitious highly viscous bosonic matter have $\gamma_{abs}\sim 10^{-4}$. Obviously, for these extreme examples the reflection coefficients $\mathcal{R}^2=1-\gamma_{abs} \simeq 1$. The conclusion is that physical matter cannot mimic the effect of full absorption as the BH membrane does, and is almost completely transparent to GWs.

The orthogonal case is represented by models with approximately full absorption, which are indistinguishable from GR BHs. Since the latter is irrelevant for the post-merger measurements we will focus on the former case, where no absorption is present, and therefore fix the reflection coefficient to one.

We stress that ultracompact objects without a horizon and that obey the equivalence principle are plausible. Examples include anisotropic stars, gravastars, and possibly other compact objects \cite{Mazur:2004fk,Pani:2015tga,Cardoso:2019rvt,Raposo:2018rjn}. These objects, under some unique circumstances, allow for such reflection properties.

To summarize, the above arguments indicate that having a partially absorbing surface is not a realistic scenario. Therefore, the absorption properties are binary: either full reflection, or complete absorption.

\subsection{Lower bound for \texorpdfstring{$\epsilon$}{}}
To derive the lower bound on epsilon shown in Fig.~\ref{fig:epsilon_trend} we first recall the formula for the optimal SNR of the signal,
\begin{gather}
    \rho^2 ~=~4\int_0^{\infty}\dfrac{|\tilde{h}(f)|^2}{S_n(f)}df~,
\end{gather}
where $\tilde{h}(f)$ is the Fourier transform of Eq.~(\ref{ht}) and $S_n(f)$ is LIGO's strain sensitivity. Since the signal is approximately monochromatic, Eq.~(\ref{freq_w}), the strain sensitivity is constant, $S_n(f) = S_n(f_R)$, where $f_R$ is the signal's frequency.
This allow us to use Parseval's theorem $\int|\tilde{h}(f)|^2df= \int|{h}(t)|^2dt$ such that the SNR becomes
\begin{gather}
    \rho^2 ~=~\dfrac{4}{S_n(f_R)}\int_0^{\infty}{|{h}(t)|^2}dt~.
\end{gather}
Next, we take the time domain waveform Eq.~(\ref{ht}) and replace the integral upper bound by some arbitrary time $T$, which corresponds to the analysis integration time.
Integration over time leads to
\begin{gather}
    \rho^2~\approx~\dfrac{\tau A^2 }{2 S_n(f_R)}\left(1-e^{-2T/\tau}\right)~.
\end{gather}
We use the amplitude from Eq.~(\ref{AE}) and assume $\tau^2\omega_R^2\gg1$,
\begin{gather}
    \rho^2~\approx~\dfrac{ 8 \Delta E }{ \omega_R^2 D_L^2  S_n(f_R)}\left(1-e^{-2T/\tau}\right)~.
\end{gather}
Finally, we extract the analysis time $T$,
\begin{gather}
    T(|\ln \epsilon|) \sim c |\ln(\epsilon)|^2 \ln\left(1 - \left(a + \dfrac{b}{|\ln(\epsilon)|} \right)^2 \right)~,
    \label{nfit}
\end{gather}
where the constants $a,b,c$ are to be determined by the numerical fit to the data points of the $90\%$ credible interval of $\log_{10} \epsilon$, see Fig.~\ref{fig:epsilon_trend}.
In general, these constants are functions of the mass, spin, strain, SNR and additional unknown systematic errors. We quantify our lack of knowledge regarding the additional errors by the constants that are determined by the fit. Providing an exact analytical expression for the constant in terms of the physical parameters requires a transfer function that includes the additional errors, nevertheless the fit to data is mostly governed by the logarithmic asymptotic behaviour which is insensitive to these changes.
Further details regarding the external effects are provided in the main text.
Eventually, the numerical fit for the data is found to be bounded from below by $\log_{10} \epsilon = {-23.7}$.
The interpretation is that, given sufficiently long analysis time, the lowest possible bound that can be measured is $\epsilon = 10^{-23.7}$.

\subsection{Data analysis details}

To analyze data spanning times $[t_0, t_1]$, we first consider a slightly longer stretch of data corresponding to $\left[t_0 - \Delta t, t_1 + \Delta t \right]$. The template is generated with duration $(t_1 - t_2) + 2 \Delta t$, starting at $t_0 - \Delta t$. Both data and template are Fourier-transformed to the frequency domain and the whitening filter is applied to both. We then transform both back to the time domain and remove the times previously added, $[t_0 - \Delta t, t_0]$ and $[t_1, t_1 + \Delta t]$, from each timeseries. We choose $\Delta t$ such that the effects of the discontinuity at the boundaries are restricted to the times we remove. The resulting timeseries' are Fourier-transformed back to the frequency domain to calculate the likelihood from the inner product of the whitened data and template.
For the damped sinusoid signal, the earlier start time is compensated in the template by increasing the initial amplitude by a factor $\exp[\Delta t / \tau]$.

We use heterodyning to minimize the computational cost of generating signal templates. The frequency-domain representation of the signal is restricted to a very narrow range around its central frequency.
This allows us to generate the time-domain signal cheaply at a low sampling frequency, and then shift the frequency-representation of this signal to the desired frequency, equivalent to generating the signal directly at a higher sampling frequency.
We first generate a time-domain damped sinusoid signal, with the desired damping time $\tau$, but at frequency $f=\SI{8}{\Hz}$.
The sampling rate is chosen to be $\SI{32}{\Hz}$ to accomodate signal components up to Nyquist-frequency \SI{16}{\Hz}, which encompasses the narrow frequency band of relevant signal content.
This signal is then Fourier-transformed to the frequency domain, using the natural frequency sampling-rate for the full duration of the signal, $(t_1 - t_0) + 2 \Delta t$.
Finally, we shift the signal to the desired frequency $f$, by placing the content of the frequency series from range $[\SI{0}{\Hz},\SI{16}{\Hz}]$ into the range $[f-\SI{8}{\Hz},f+\SI{8}{\Hz}]$.
The resulting frequency domain waveform is then used for the likelihood calculation.

For long analysis durations, the Doppler shift due to the orbital motion of the Earth becomes time-dependent. However, we find this to be negligible for the durations of less than $\sim \SI{1000}{\second}$ used in this analysis, and consider only a static Doppler shift.

\subsection{Validation with noise and simulated signals}

To validate our results we repeat our analysis on off-source detector noise (before GW150914) and on simulated Gaussian noise. These serve to determine the analysis' diagnostic power when no signal is present in the noise. We also add simulated signals to both the off-source data to verify the effectiveness of the analysis to detect known signals.

In each case we analyse 128 seconds of data for the presence of a signal and use 512 seconds of data before the analysis window to estimate the power spectral density (PSD). For the Gaussian noise case, the noise is coloured to agree with the PSD estimated from off-source data at times before GW150914.

\begin{figure}
    \includegraphics[width=\columnwidth]{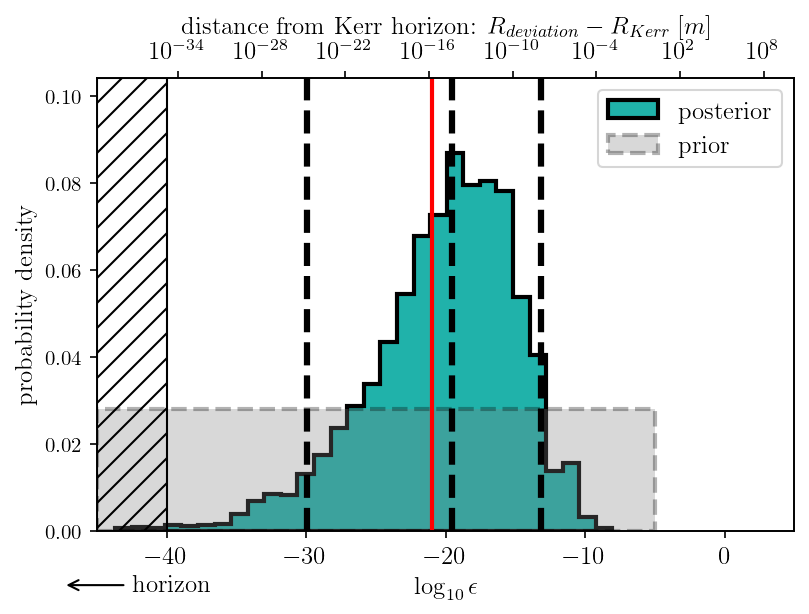}
    \caption{Same as Fig.~\ref{fig:epsilon_posterior} with a simulated signal injected into detector noise.
        The histogram shows the marginal posterior for $\epsilon$, the shaded region is the prior.
        The red line marks the value for $\log_{10} \epsilon$ of the simulated signal.
        The dashed lines indicate the (two-sided) 90\% credible interval and the median value, respectively.
        The posterior clearly prefers non-zero values of $\epsilon$ in the presence of the simulated signal, and the recovered value for $\log_{10} \epsilon$ is within the 90\% credible interval.
    }
    \label{fig:epsilon_posterior_injection}
\end{figure}

For the off-source real detector noise analysis, we find that the source-parameter posteriors are unchanged from their priors. The posteriors for $\log_{10} \epsilon$, $\tau$ and $A$ are consistent with the expectation for noise without a signal. Smaller $\epsilon$ corresponds to smaller signal amplitudes and longer damping times, as the same total energy is radiated away over increasingly long times. We find the posteriors prefer large $\tau$ and small $A$ and $\epsilon$, with the latter peaking at the lower prior boundary. The one-sided 90\% credible interval bound for $\log_{10} \epsilon$ is $-23.7$.

The frequency posterior shows narrow peaks for specific frequencies, often associated with increased SNRs. These peaks appear only for long analysis durations and become more dominant with increasing duration. We can attribute the most prominent peaks to lines in the power spectral density of the noise, such as the \SI{180}{\Hz} harmonic of the \SI{60}{\Hz} line resulting from the AC power grid frequency.

The simulated Gaussian noise analysis yields similar results as the off-source detector noise case, with the source-parameter posteriors unchanged from their priors. Large $\tau$ and small $A$ and $\epsilon$ are preferred, with $\epsilon$ peaking at the lower prior boundary, and the 90\% bound being $\log_{10} \epsilon = -22.9$.

Both cases show the narrow peaks in the frequency posterior described before.
The peaks are more pronounced for real detector noise than for simulated Gaussian noise.
The most prominent peaks coincide with lines of excess power in the PSD for the detector noise, but not for simulated Gaussian noise colored with the same PSD.
This suggests the presence of non-Gaussian noise features in the real noise that are partially matched by the sinusoidal templates.
Slow variations of the PSD in the detector noise may amplify this effect.
For the analysis, the PSD has to be estimated from off-source data, such that slow variations in the line parameters cannot be corrected for in long-duration analyses.

We perform several analyses with simulated signals added to the off-source noise, for example with $\log_{10} \epsilon = -18$ or $\log_{10} \epsilon = -21$.
For each simulation, the injected value lies within the 90\% credible interval of the $\epsilon$-posterior, and the posterior peaks away from the lower prior boundary and near the correct value.
Figure~\ref{fig:epsilon_posterior_injection} shows this for the $\log_{10} \epsilon = -21$ injection.
For all injections, the frequency posterior is concentrated in a narrow peak around the correct frequency, limited by the frequency-resolution of the data.
The one-sided 90\% bounds for these injections into detector noise is $\log_{10} \epsilon = -13.3$ and $\log_{10} \epsilon = -14.3$, respectively, larger than found for noise without a signal.
As we are expecting a signal in the injection case, we also use the two-sided credible interval as shown in Figure~\ref{fig:epsilon_posterior_injection}.
The ranges recovered then are $\log_{10} \epsilon = -18.3^{+6.3}_{-10.0}$ and $\log_{10} \epsilon = 19.5^{+6.4}_{-10.4}$ for the louder and quieter injection, respectively.

\end{document}